# Communication

## "Perfect Echo" HMQC: Sensitivity and Resolution Enhancement by Broadband Homonuclear Decoupling

Bikash Baishya[a]*, C. L Khetrapal[a], Krishna Kishor Dey[a]*

[a]*Center of Biomedical Magnetic Resonance, SGPGIMS Campus, Raebareli Road, Lucknow, 226014, India*

**ABSTRACT**

Homonuclear $^1$H-$^1$H *J*-modulation leads to *J*-multiplets in $F_1$ dimension of 2D $^1$H-$^{13}$C HMQC spectra. This hampers unambiguous signal assignment for overcrowded $^{13}$C spectra. Broadband homonuclear decoupling has been achieved in the indirect $t_1$ evolution period by incorporating blocks of perfect echo. This method enhances resolution and sensitivity of 2D $^1$H-$^{13}$C HMQC spectra. The results on Cyclosporine demonstrate that the method is very efficient for -$^{13}$CH$_2$ groups, though partial sensitivity and resolution enhancements have also been observed for -$^{13}$CH and -$^{13}$CH$_3$ groups. Interpretation of the result based on product operator formalism is also given.

**Introduction:**

The spin echo [1] block has been extensively used to design experiments to suppress the effects of chemical shifts and field inhomogeneity, to measure spin-spin relaxation rate ($R_2$), to estimate translational diffusion coefficient and to address chemical exchange. In addition, the block is routinely applied for designing multidimensional NMR experiments. However, the ubiquitous presence of homonuclear *J*-couplings in spin echo can restrict their use in many situations. Homonuclear *J*-coupling being a bilinear interaction, their broadband refocusing is not possible. In order to overcome this problem, many techniques have been developed such as resonance-specific selective irradiation [2, 3] and a variety of 1D [4-19] and 2D approaches [20-35].

* Corresponding author. Fax: 91-522-2668215
*E-mail address:* bikashbaishya@gmail.com (Bikash B)
dey.krishna@gmail.com (Krishna K D)

Noteworthy among the 1D approaches are, rapid refocusing (or spin lock) in the *Carr-Purcell-Meiboom-Gill* (CPMG) experiment [4-6] and, slow refocusing in conjunction with cumulative pulse imperfections at favourable resonance offsets [7-12]. Other 1D schemes are "perfect echo" approach for an weakly coupled AX spin system [13, 14] and their recent extension to arbitrary spin systems [15], zero/double quantum filtration methods in two spin systems [16-17] and, homonuclear decoupling in a continuous 1D scan by supercycling BIRD (Bilinear rotation decoupling) blocks [18, 19]. Worthwhile to mention among the 2D approaches are 2D *J*-spectroscopy [20], the use of BIRD pulses [19, 22], and use of spatially-selective $^1$H spin inversions in Zangger/Sterk sequence [30]. All the 2D approaches offer high spectral resolution albeit at the cost of a relatively lengthy multi-$t_1$ acquisitions irrespective of sensitivity considerations. Techniques have also been developed to reduce such lengthy 2D acquisition time [33, 35].

While homonuclear J-coupling is a rich source of structural information in liquid state NMR, its presence in certain circumstances can degrade the performance of 2D heteronuclear multiple quantum spectroscopy [36-41]. The heteronuclear multiple quantum coherence in the indirect $t_1$ dimension evolves under the passive $^1$H-$^1$H *J*-couplings leading to unresolved multiplets and phase distortions. This can hamper unambiguous signal assignment for overcrowded spectra in the indirect domain frequency axis ($F_1$). The unresolved multiplets can further reduce sensitivity. To circumvent this problem, spin-locked multiple quantum coherence for signal enhancement in heteronuclear multidimensional NMR has been developed [42]. Weak spin lock on H$^\alpha$ is applied during a constant-time period in 2D HMQC experiment to remove the $^1$H-$^1$H *J*-dephasing focusing mainly on methine (-CH) groups in proteins. Potential drawbacks mentioned are: off-resonance effects introducing residual $J(^{13}C^\alpha$-H$^\alpha)$ modulation and Hartmann-Hahn transfer for small difference in H$^\alpha$-H$^\beta$ chemical shift. Very high resolution studies will demand long $t_1$ spin lock and therefore, would be less preferable. In addition constant-time approach displays lower sensitivity by virtue of the longer duration of the sequence for all values of $t_1$. Band selective, constant-time

HMBC has also been reported to enhance the $F_1$ resolution by suppressing the unwelcome proton *J*-coupling modulation [43-44]. Recently, 2D tilt HMBC has also been proposed that produces broadband homodecoupled HMBC spectra recorded as 3D and without resorting to constant time approach [45]. Thus, there is scope for exploring other alternative homonuclear decoupling in such circumstances. Therefore, in the present study we exploit perfect echo based homonuclear decoupling in the indirect $t_1$ dimension of 2D HMQC. A perfect echo sequence is essentially a double spin echo with insertion of a 90° pulse at the centre of the two refocusing pulses. We demonstrate that perfect echo based broadband homonuclear decoupling enhances the sensitivity and resolution of 2D HMQC. In addition the method allows flexibility to set up such experiment in a non constant-time manner incorporating just a few more extra pulses.

**Description of the 'perfect echo' HMQC pulse sequence:**

The modified HMQC pulse sequence called 2D perfect echo HMQC (abbreviated pe-HMQC) with n-perfect echo block spanning the $t_1$ evolution period shown in Fig. 1(b) is modified from phase sensitive gradient selection-HMQC [46]. This sequence involves coherence selection after the $t_1$ period and is ideally suited for our purpose. For n=1, i.e. one pe block during $t_1$, the refocusing pulse at the centre of the $t_1$ evolution period of conventional HMQC is replaced by a 90° pulse and at the centre of each $t_1/2$ period a refocusing pulse is introduced. For n=2, two perfect echo blocks span the $t_1$ period. Only the $I^+S^+$ heteronuclear multiple quantum coherence evolving during $t_1$ contributes to the final signal during acquisition. Here, *I* corresponds to $^1$H spin operator and *S* to $^{13}$C spin operator. In a conventional HMQC pulse sequence, the $I^+S^+$ multiple quantum coherence does not evolve under the $^{13}$C-$^1$H $^1J$ – couplings and $^{13}$C-$^1$H $^nJ$ -couplings (where n>1) are refocused by the refocusing pulse on protons at the centre of $t_1$ evolution period. However, $^1$H-$^1$H passive *J*-couplings being bilinear continue to modulate the multiple quantum coherence during $t_1$. Considering n=1 pe-HMQC, the 1$^{st}$ 180° pulse on protons in the perfect echo block refocuses proton chemical shifts and $^nJ(^{13}$C-$^1$H) (where n>1) evolutions exactly at the centre of the bloch at time point 'b'. Subsequently a 90° pulse is applied, which exchanges antiphase magnetizations of coupled proton spins. Consequently, in-phase magnetisation is restored at the end of the second echo at time point 'd' produced by the 2nd $^1$H refocusing pulse of the perfect echo block [13,14].

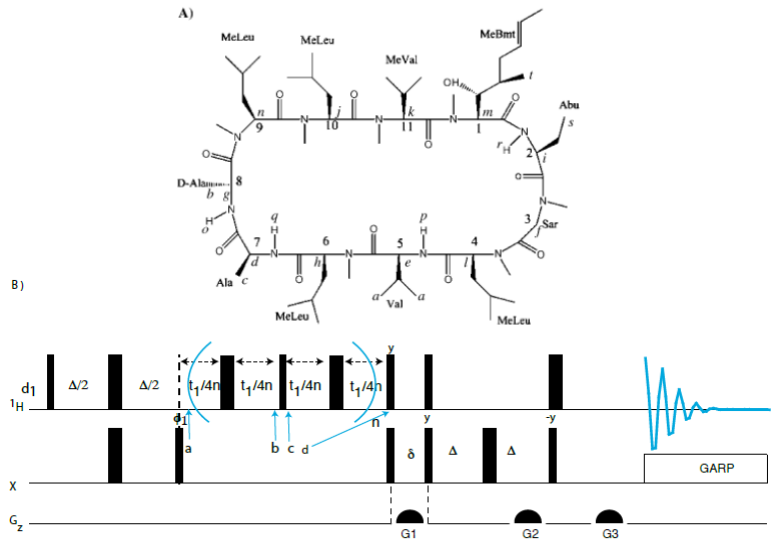

**Fig. 1. (a)** Structure of Cyclosporine-A with its amino acids numbered 1 to 11 **(b)** 2D pe-HMQC pulse sequence with n-perfect echo block during $t_1$ domain. The experimental results presented in this paper are obtained with n=1. Thin pulses are 90° and thick pulses are 180°. All pulses are x phase unless marked with a different phase. All gradients are along z axis. G1=17 and is a purge gradient. Coherence selection is carried out by G2:G3= 50:55. $\phi_1$=x,-x and Receiver phase x, -x. States TPPI for frequency discrimination in $t_1$ and Garp decoupling during acquisition is used.

The product operator formalism is given below for pe-HMQC (with n=1) to account for the sensitivity enhancement for an isolated methylene group where the two geminal protons are weakly coupled i.e. $^1H_a$-$^{13}C$-$^1H_b$.

The operators at time point 'a': $2I_{1X}S_Y + 2I_{2X}S_Y$

Where $I_1$ is $^1H_a$ and $I_2$ is $^1H_b$. Note equal initial $I_X S_Y$ magnetisation is created for both $^1H_a$-$^{13}C$ and $^1H_b$-$^{13}C$ pair as the two $^1J(^{13}C$-$^1H)$ couplings are equal.

The operators at time point b: $2S_Y \cos\left(\Omega_S \frac{t_1}{2}\right)[A] - 2S_X \sin\left(\Omega_S \frac{t_1}{2}\right)[A]$

Where $A = \left[I_{1X}\cos\left(\pi J_{IK}\frac{t_1}{2}\right) + 2I_{1Y}I_{2Z}\sin\left(\pi J_{IK}\frac{t_1}{2}\right) + I_{2X}\cos\left(\pi J_{IK}\frac{t_1}{2}\right) + 2I_{2Y}I_{1Z}\sin\left(\pi J_{IK}\frac{t_1}{2}\right)\right]$

The operators at time point 'c' just after the J-refocusing 90°x pulse can be written as: $2S_Y \cos\left(\Omega_S \frac{t_1}{2}\right)[B] - 2S_X \sin\left(\Omega_S \frac{t_1}{2}\right)[B]$

Where B $B = \left[I_{1X} \cos\left(\pi J_{IK} \frac{t_1}{2}\right) - 2I_{1Y}I_{2Z} \sin\left(\pi J_{IK} \frac{t_1}{2}\right) + I_{2X} \cos\left(\pi J_{IK} \frac{t_1}{2}\right) - 2I_{2Y}I_{1Z} \sin\left(\pi J_{IK} \frac{t_1}{2}\right)\right]$

Thus exchange of antiphase states by J-refocusing 90°x pulse is evident

During the next half of $t_1/2$ evolution period from 'c' to 'd' the in-phase operator can be restored as detailed in [14] and the operator at time point 'd' will be:

$$2I_{1X}S_Y \cos(\Omega_S t_1) + 2I_{2X}S_Y \cos(\Omega_S t_1)$$

These terms are free from $^1$H-$^1$H J-modulations [do not contain the term $\cos(\pi J_{IK} t_1)$] and continue the normal course of the later part of HMQC sequence. The product operator approach shows that J-refocusing is very efficient for -CH$_2$ and -NH$_2$ groups where the geminal protons are weakly coupled as will be demonstrated in experimental section as well. It can be mentioned here, that, the BIRD method of decoupling cannot be implemented during $t_1$ of HMQC as it refocuses the $^{13}$C chemical shift evolution and particularly for -CH$_2$ groups the BIRD decoupling does not work.

The sensitivity loss during a conventional 2D HMQC experiment can be explained by considering the operators present at time point 'b'. Firstly, the antiphase part $2I_{1Y}I_{2Z}$ and $2I_{2Y}I_{1Z}$ continues to evolve instead of refocusing as the J-refocusing 90° pulse is absent. These terms lead to superposition of in phase absorptive and antiphase dispersive lineshape in F$_2$. These antiphase terms are removed either by a 90°$_y$ purge pulse just before acquisition [48] or by a killer gradient G$_1$ as in the pulse sequence of Fig. 1(b) executed without the pe block. Secondly the term $2I_X S_Y \cos(\Omega_S t_1)\cos(\pi J_{IK} t_1)$ displays a doublet along F$_1$. In pe-HMQC due to refocusing of the homonuclear J-coupling interactions the antiphase part can be refocused to inphase signal and the doublet collapses to a singlet leading to gain in sensitivity and line narrowing. Note the final term at time point 'd' $2I_{1X}S_Y \cos(\Omega_S t_1) + 2I_{2X}S_Y \cos(\Omega_S t_1)$ does not contain $\cos(\pi J_{IK} t_1)$ i.e. a singlet instead of a doublet will be observed.

The exchange of antiphase $^1$H magnetisation at time point 'c' does not give rise to any unwanted $^{13}$C-$^1$H cross peak as long as $^1$H chemical shift is refocused during entire $t_1$ evolution. Provided $2\pi J\tau \ll 1$ and we start with equal initial magnetisations for the two coupled spins the refocusing of homonuclear J-couplings is efficient [13-15].

The $^1$H-$^1$H J-refocusing by pe-HMQC seems to be less efficient in a spin system of the form $^{13}$C($^1$H)-$^{12}$C($^1$H) which is the conventional spin system such heteronuclear experiment selects as it deals with 1.1 % isotopomers of $^{13}$C in natural abundance. The $^1J(^{13}C$-$^1H)$ (140-160Hz) and $^2J(^{13}C$-$^1H)$ (2-15Hz) being very different in magnitude, equal initial magnetisation is difficult to create at time point 'a'. In such a situation, the starting operators at time point 'a' of the pulse sequence in Fig. 1(b) will be of the form $2I_{1X}S_Y + f2I_{2X}S_Y$ where $I_1$ operator corresponds to $^1$H bound to $^{13}$C, and $I_2$ corresponds to $^1$H bound to $^{12}$C and $S$ operator is for $^{13}$C. The magnitude of factor $f$ is equal to $\sin[\pi(^2J_{CH})2\tau]$, depends on the magnitude of $^2J(^{13}C$-$^1H)$ couplings and the length of the INEPT transfer delays $2\tau$ and is quite small compared to the 1$^{st}$ term. This 2$^{nd}$ term with factor $f$ represents the extent of J-refocusing in case of pe-HMQC compared to conventional HMQC without pe block for such spin systems. For $^2J$=15Hz and $2\tau = 3.3$ ms the value of $\sin(\pi\,^2J_{CH}2\tau)$ is 0.16. In case there are more protons at neighbouring $^{12}$C sites then the contribution from too many $f$ factors can still enhance the sensitivity.

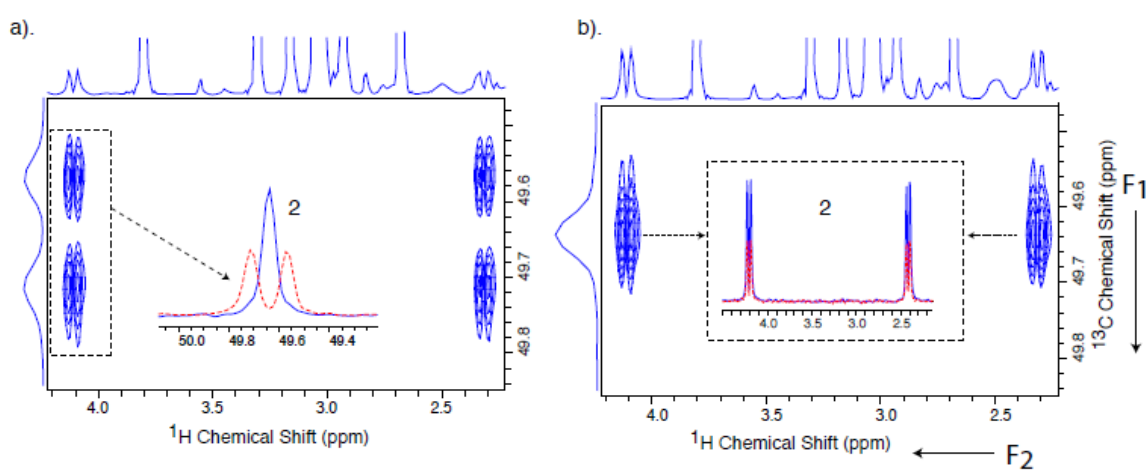

**Fig. 2.** (a) The methylene cross peaks of conventional $^1$H-$^{13}$C 2D HMQC spectrum for the Sar-3 residue of CsA along with the corresponding projections. The 2D data matrix is 840 and 2400 points in F$_2$ (SQ) and F$_1$ (3Q)

dimensions respectively. Spectral widths of 2790 Hz and 6500 Hz were chosen in the direct and indirect dimensions respectively. The number of accumulations was 4 for each $t_1$ increment. Relaxation delay used was 1.8 sec. The time domain data was processed by zero filling it to 1 k and 4k points in $F_2$ and $F_1$ dimensions respectively, with sine square bell window function. The spectrum is displayed after phase correction with a spectral resolutions of 2.72Hz and 1.58Hz in the direct and indirect dimensions respectively. The multiplets along $F_1$ axis are clearly visible. **(b)** The same spectral region as in (a) obtained by $^1$H-$^{13}$C pe-HMQC sequence along with the corresponding projections. Both (a) and (b) are plotted on the same contour level with same experimental and processing parameters. Collapse of the doublets to singlets along $F_1$ is noteworthy as compared to $F_1$ axis of (a). In 2(a) the dotted arrow from the box displays the $F_1$ cross section taken for that cross peak at maximum value, in dotted line conventional HMQC and in continuous line pe-HMQC. The dotted box inside Fig. 2b, together with the arrows displays the $F_2$ cross section taken for the two cross peak, in dotted conventional HMQC and in continuous pe-HMQC.

It can be mentioned here that the final in-phase spin state obtained is the result of evolutions of different states with different relaxation rates. Mixing of relaxation rates in addition to polarization transfer occurs and is more significant for higher spin order [14]. Because of different $t_1$ increments in the pe-HMQC sequence, the echo time does not remain constant for the perfect echo. The build up of $^1$H antiphase states can be significant for very long $t_1$ evolution period and their comparatively faster decay could hamper effective refocusing of $^1$H-$^1$H *J*-dephasing. However, this issue is not as severe as gain in in-phase intensity over a long range of echo times is demonstrated in literature [14, 15].

**Experimentals:**

In order to compare the sensitivity and resolution of 2D pe-HMQC pulse sequence with conventional 2D HMQC sequence, $^1$H-$^{13}$C 2D HMQC variant of the sequence was performed on Cyclosporine sample (50mM in deuterated benzene, $C_6D_6$) on a Bruker avance 400 MHz NMR spectrometer equipped with a 5 mm BBI probe. The chemical structure of Cyclosporine-A is shown in Fig.

1(a). Detail of experimental and processing parameters are given in figure captions.

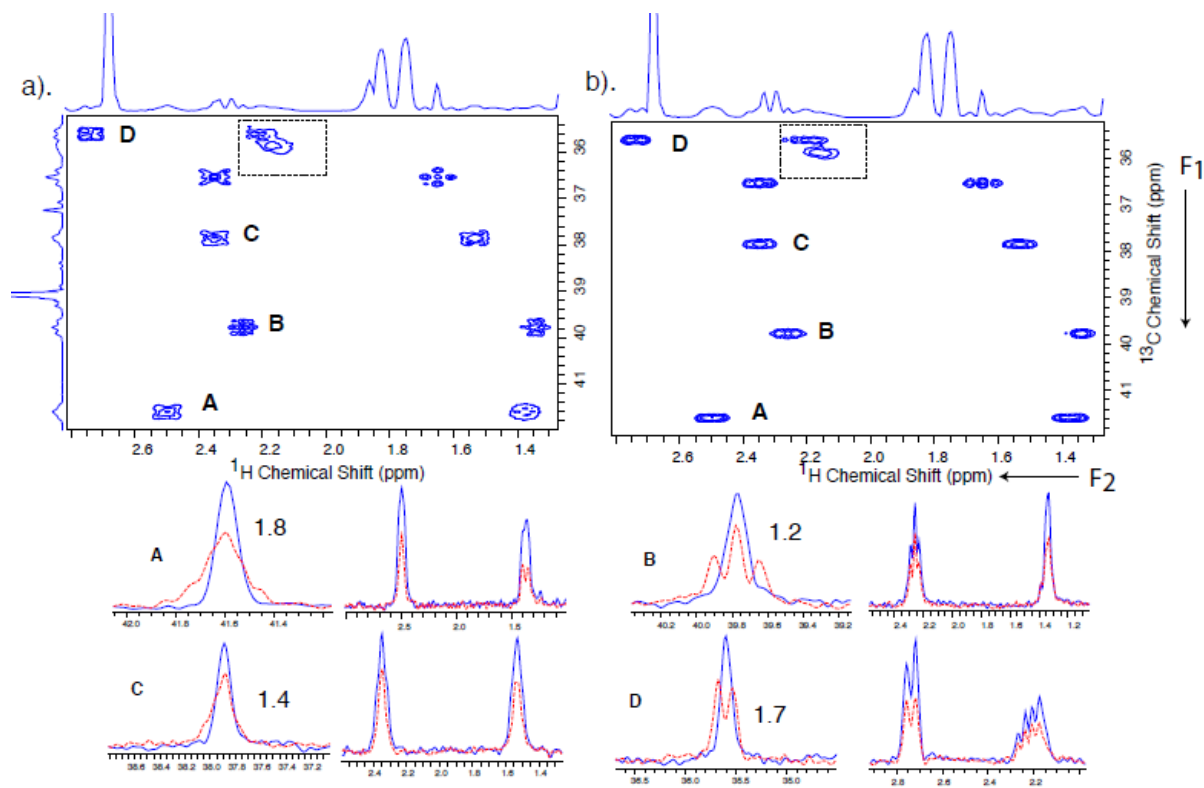

**Fig. 3. (a)** The methylene region of conventional $^1$H-$^{13}$C 2D HMQC spectrum of CsA along with the corresponding projections. All the experimental and processing parameters are kept same as in Fig. 2. The multiplets for the cross peak along $F_1$ axis are clearly visible. **(b)** $^1$H-$^{13}$C pe-HMQC spectrum of the same region as in (a) along with the corresponding projections plotted on the same contour levels. All the experimental and processing parameters are kept same. Collapse of the multiplets along $F_1$ is noteworthy as compared to $F_1$ axis of (a). Distortion of the cross peaks along $F_1$ is considerably less. $F_1$ and $F_2$ cross sections at maximum value of the cross peaks are extracted for peaks A to E in both spectra (a) and (b) and comparison is displayed below. Continuous line corresponds to pe-HMQC while dotted line corresponds to conventional HMQC without pe block. The gain in signal to noise ratio by pe-HMQC is reported by the side of the peaks. Comparison of the dotted box inside (a) and (b) reveals the two closely resonating carbon peaks are better resolved in (b).

**Results and discussions:**

As detailed in the theoretical consideration, *J*-refocusing is most efficient in an isolated -$CH_2$ groups, therefore, we focus our attention initially on such a group in Sar-3 residue of CsA displayed in Fig. 1(a) marked 'f'. There are two -N-$CH_3$ groups which are at four and five bonds apart and therefore the long range $^1H$-$^1H$ *J*-couplings can be safely ignored. Fig. 2(a) displays the methylene cross peaks of Sar-3 in conventional $^1H$-$^{13}C$ HMQC spectrum along with the corresponding projections. Since there are two weakly coupled protons correlating to the same $^{13}C$, two cross peaks for the same $^{13}C$ resonance are observed. The $^1H$-$^1H$ $^3J$- modulations (13.9Hz) gives rise to doublet in both $F_1$ and $F_2$ axis. Fig. 2(b) displays the $^1H$-$^{13}C$ pe-HMQC spectrum with one pe block in $t_1$ domain, where $F_1$ cross section collapses to a singlet. In Fig. 2a the dotted arrow from the box displays the $F_1$ cross section taken for that cross peak, in dotted line conventional HMQC and in continuous line pe-HMQC. We measured a gain in signal to noise ratio by a factor of two for this cross peak reported by the side of the peak. The dotted box inside Fig. 2b, together with the arrows displays the $F_2$ cross section taken for the two cross peaks, in dotted line conventional HMQC and in continuous line pe-HMQC. The same gain in signal to noise ratio is evident. The $t_1^{max}$ and $t_1^{min}$ in the experiment was 369.2ms and 186us respectively with 2400 $t_1$ increments. This led to $\tau$ values in the range from 46.5us to 92.3ms and even for longer $\tau$ values the decoupling efficiency is retained.

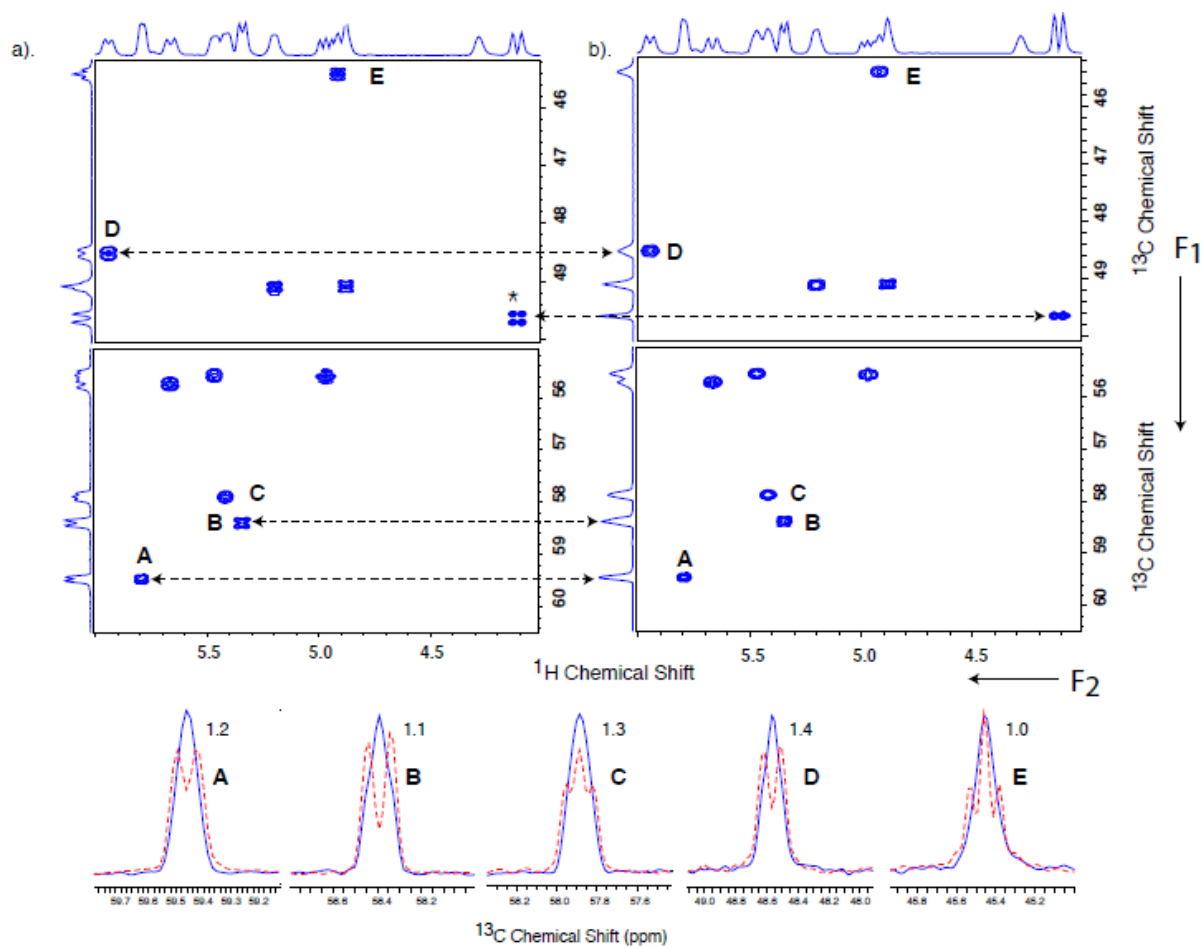

**Fig. 4. (a)** The methine region of conventional $^1$H-$^{13}$C 2D HMQC spectrum of CsA along with the corresponding projections. The partially resolved multiplets along F$_1$ axis are clearly visible. **(b)** $^1$H-$^{13}$C pe-HMQC spectrum of the same region as in (a) along with the corresponding projections plotted on the same contour levels. The experimental and processing parameters are same for both (a) and (b). Collapse of the multiplets along F$_1$ is noteworthy as compared to F$_1$ axis of (a). F$_1$ and F$_2$ cross sections are extracted at maximum value of the cross peaks for peaks A to E and comparison is displayed below, where continuous line corresponds to pe-HMQC and dotted line corresponds to conventional HMQC without pe block. The gain in signal to noise ratio by pe-HMQC sequence compared to conventional HMQC sequence is reported by the side of the peaks.

Fig.3 displays a few other methylene region cross peaks of CsA in conventional $^1$H-$^{13}$C HMQC spectrum (a) and pe-HMQC spectrum (b) along with the corresponding projections. This region is part of the same spectrum as shown in Fig. 2 but plotted separately for clarity. All the experimental and processing parameters are kept same as in Fig 2. Both (a) and (b) are plotted on the same contour level. Collapse of the multiplet pattern of the cross peaks to singlet along $F_1$ axis of (b) is noteworthy as compared to (a) where unresolved multiplets along $F_1$ is visible for all cross peaks. $F_1$ and $F_2$ cross sections are extracted for peaks A to E at their maximum peak height from both spectra and comparison is displayed below, where continuous line corresponds to pe-HMQC while dotted line corresponds to conventional HMQC without pe block. In A to E below, left peak $F_1$ cross section and right peaks $F_2$ cross section. The gain in signal to noise ratio by pe-HMQC compared to conventional HMQC is reported by the side of the peaks. Comparison of the dotted box between (a) and (b) reveals the two closely resonating carbon peaks are better resolved in (b).

Fig. 4 displays the methine (-CH) region cross peaks of CsA in conventional $^1$H-$^{13}$C HMQC spectrum (a) and pe-HMQC spectrum (b) along with the corresponding projections. This region is part of the same spectrum as in Fig. 2 and 3 and therefore, pertains to same experimental and processing parameters. The multiplets along $F_1$ projection of (a) are clearly visible and these multiplets collapse to singlet in pe-HMQC $F_1$ projection shown in (b). $F_1$ and $F_2$ cross sections are extracted for peaks A to E for both spectra and comparison is displayed below for $F_1$ where continuous line corresponds to pe-HMQC and dotted line corresponds to conventional HMQC without pe block. The partial collapse of the multiplets is visible along $F_1$. Gain in signal to noise ratio by pe-HMQC sequence compared to conventional HMQC sequence is reported by the side of the peaks. Thus, even for –CH region cross peaks, *J*-refocusing and sensitivity gain is observed.

Partial sensitivity enhancement and collapse of the multiplets were observed even for the methyl region of CsA in pe-HMQC.

In conclusion we have demonstrated homonuclear decoupling in the indirect dimension of 2D HMQC experiment. This decoupling works very efficiently for -$^{13}$CH$_2$ groups where the two protons are weakly coupled. Significant line narrowing and sensitivity enhancement is observed in –CH2 region. Two closely resonating $^{13}$C in this region could be resolved better. The method demonstrates partial sensitivity enhancement and collapse of the multiplets for many other type of peaks (-CH, -CH$_3$, etc.). pe-HMQC will be useful for high resolution studies of overcrowded spectra of natural products and biomolecules.

**Acknowledgement:** We acknowledge Science and Engineering Research Board (SERB), India for financial support provided through grant no. SR/FT/CS-123/2011.